\documentstyle[11pt,newpasp,twoside,epsf,psfig]{article}
\def\edcomment#1{\iffalse\marginpar{\raggedright\sl#1\/}\else\relax\fi}
\reversemarginpar

\begin{document}
\title{HI Signatures of Reionization}

\author{Paolo Tozzi} 

\affil{Oss. Astronomico di Trieste, via G.~B. Tiepolo 11,
34100 Trieste, Italy}  

\begin{abstract}
The exploration of the end of the Dark Ages will be one of the most
exciting field of the next decade.  While most of the proposed
observations must await the next--generation telescopes, the
observational window of the redshifted 21cm line offers the
possibility to investigate the physics of reheating and reionization
on a short term.  Here we describe several possible signatures
detectable in the wavelength range 100-200 MHz.  Among the physics
that can be investigated: the epoch of reheating and reionization;
topology and timescales of reheating; the nature of the ionizing
sources; the baryon distribution at redshift $z\sim 10$.  Such a good
deal of information is within reach of present--day, or near--future
radio facilities.
 
\end{abstract}

\section{Exploring the Dark Ages}

The Dark Ages are ended by the appearance of the first stars and/or
quasars, that reheat the diffuse cosmic baryons and then reionize the
Universe (see Rees 1999).  The reionization is defined as the epoch
when the volume--filling factor of HII regions is $\simeq 1$, and the
Universe becomes transparent to ionizing radiation (for an extensive
review see Loeb \& Barkana 2001).  The absence of a Gunn--Peterson
(GP) effect in the line of sight of distant quasars, put the epoch of
reionization ($z_{reion}$) at redshifts $>5$.  To date, there are
little additional constraints on the physics of the cosmic baryons at
such high redshifts.  This is also due to the complexities involved by
any theoretical model that must include the nature and the birthrate
of the first luminous objects, their spectrum and emissivity, their
feedback into the surrounding Intergalactic Medium (IGM), etc.

A significant step forward has been recently obtained with Keck
spectroscopy of the most distant SLOAN quasars (Becker et al. 2001;
Djorgovski et al. 2001).  The presence of a sudden increase in the
Ly$\alpha$ opacity between redshift 5 and 6 may indicate that the
Universe is approaching reionization at $z\sim 6$.  However, Barkana
(2001) pointed out that, due to the high Ly$\alpha$ opacity, these
observations are also consistent with a post--reionization phase, and
that a conclusive proof requires the observation of similar GP
absorptions along several additional line of sights.  Alternatively,
the observation of smoking--gun signatures has been proposed.  Among
them: dectection of scattered Ly$\alpha$ emission around sources
beyond $z_{reion}$ (Loeb \& Rybicki 1999); features in galaxy number
counts at $z>z_{reion}$ (Barkana \& Loeb 2000).  These observations
would be effective probes of the physics of cosmic baryons and of the
nature of the first ionizing sources at the same time, but they must
wait for NGST and/or SIRTF, which will be operating at the end of this
decade.  Signatures of the reionization are expected also in the CMB.
MAP and Planck can probe the optical thickness at $z_{reion}$ in the
30--150 GHz range.  In particular, fluctuations on $\le 0.1$ degree
reflect the topology of reionization.  However, such measures are
highly degenerate with cosmological parameters.

It is clear that the investigation of the Universe at $z\simeq 10$ is
a difficult task with present--day facilities.  In this situation, the
observational window offered by the redshifted 21cm line can open much
of the Universe to a direct study of the reheating and reionization
epochs.  In the following we will discuss some observations of HI at
high $z$ that can shed light on many aspects of these processes.

\section{The History of the Cosmic Baryons}

In this section we outline a simple scenario for the thermal history
of the diffuse cosmic baryons in the high--$z$ Universe, and we show
under which condition they emit in the 21cm line.

During the Dark Ages ($z>10$), most of the baryons of the Universe are
distributed uniformly.  The typical density contrast on scales $> 1\,
h^{-1}$ Mpc is still in the linear regime.  The temperature of the
diffuse baryons, experiencing only adiabatic cooling since the
decoupling from the cosmic background radiation, is about $T_K = 0.026
\times (1+z)^2$ K (Couchman 1985), which, at redshift $z\simeq 10$, is
one order of magnitude smaller than the CMB temperature $T_{CMB} =
2.73\times (1+z)$.  The spin temperature $T_S$ of the cosmic baryon is
defined via the ratio of the population of the triplet and the singlet
state of the spin transition originating the 21cm emission.  In
absence of any coupling, the spin of the IGM is in thermal equilibrium
with the background radiation, and $T_S=T_{CMB}$ on a very short
timescale.  To observe the 21cm emission from HI against the CMB, we
need some process that decouple $T_S$ from $T_{CMB}$.  Collisions are
ineffective, since the density contrast on Mpc scales is too low
(Madau, Meiksin \& Rees 1997).

There is, however, an efficient mechanism that makes HI visible in the
redshifted 21cm line: the Wouthuysen--Field effect (WT).  In this
process, a Ly$\alpha$ photon field mixes the hyperfine levels of
HI in its ground state via intermediate transition to
the $2p$ state (see Appendix in Tozzi et al. 2000).  The process
effectively couples $T_S$ to the color temperature $T_\alpha$ of a
given Ly$\alpha$ radiation field (Field 1958).  The color temperature
is easily driven toward the kinetic temperature $T_K$ of the diffuse
IGM due to the large cross section for resonant scattering (Field
1959).  Therefore the spin temperature can be written as:
\begin{equation}
T_S=\frac{T_{\rm CMB}+y_\alpha T_K}{1+y_\alpha}\, ,
\label{eq1}
\end{equation}
where $y_\alpha\simeq 3.6\times  10^{13}P_\alpha/T_K$ is the Ly$\alpha$
pumping efficiency, and $P_\alpha$ is the total rate at which
Ly$\alpha$ photons are scattered by an hydrogen atom.

Thus, the diffuse IGM at very high redshift (between recombination and
full reionization) can be observed in the redshifted 21cm radiation
against the cosmic background in presence of a Ly$\alpha$ background
(likely produced by the same sources that reionize the Universe).  The
expected emission is less than 10 mK (Shaver et al. 1999; Tozzi et
al. 2000).  However if $T_S \simeq T_K \le T_{CMB}$, the HI is visible
in absorption with a signal significantly larger than the maximum
detectable in emission (see Hogan \& Rees 1979).  Thus, the most
favourable condition for detection is given by a cold IGM in a strong
Ly$\alpha$ field.

At the same time, the Ly$\alpha$ photon field also reheats the diffuse
gas, driving $T_K$ towards larger values.  The thermal history of the
diffuse IGM results from the competition between adiabatic cooling and
reheating due to the photon field, and it can be simply described as:
\begin{equation}
{{dT_K}\over{dz}}={{2\mu}\over 3}{\dot E\over k_B}
{{dt}\over{dz}} + 2{{T_K}\over{(1+z)}},
\label{eq2}
\end{equation}
where $\dot E$ is the heating rate due to recoil of scattered
Ly$\alpha$ photons and, possibly, to additional contribution from an
X--ray background (here $\mu=16/13$).  The heating is quite efficient,
since only 0.004 eV per particle are needed to heat the IGM above the
CMB at $z\simeq 10$.  Therefore, Ly$\alpha$ resonant scattering or
photoelectric heating by soft X--ray photons can reheat the IGM in
10\% of the Hubble time (Madau, Meiksin \& Rees 1997), and the WT
mechanism leads preferentially to emission in the 21cm line.

The signal from high--$z$ HI can be predicted by solving Eq. 1 and
Eq. 2, after assuming a value for the Ly$\alpha$ radiation field.  A
reference value is the thermalization rate $P_{th} = (7.6 \times
10^{-12} \, s^{-1})(1+z)/10$, which is the critical value that would
drive $T_S$ towards $T_K$, and, at the same time, would heat the
baryons at a rate of about 200 K per Gyr (see Madau, Meiksin \& Rees
1997).

\section{21cm Signatures of Reionization}

Between $5<z_{reion}<20$ the Universe is expected to experience a
phase transition from the neutral to the reionized phase.  A first
effect of the reionization is the sudden disappearance of the 21cm
emission from HI.  In fact, Shaver et al. (1999) showed that with a
bandwidth of 5 MHz, and an observation of 24 hours, the reionization
signal (Figure 1) is detected at 100$\sigma$ independent of telescope
size.  For this measure sensitivity is not an issue, the challenge
concerns signal contamination.  In fact, the sharpness of this
feature, expected in the 70--240 MHz range, makes it recognizable from
galactic and extragalactic foregrounds (Di Matteo et al. 2001).  In
addition, Gnedin \& Ostriker (1997) computed the 21cm emission in a
specific $\Lambda$CDM cosmology within an N--body simulation, and
showed that the fluctuations in the 21cm emission will dominate the
fluctuations in the CMB by two orders of magnitude (Figure 2).  The
global, all--sky signature of the reionization, can be easily detected
by present--day radio telescopes also in the case of a
non-homogeneous, patchy process, as in the case of isolated, strong
ionizing sources.

\begin{figure}
\centering
\plotone{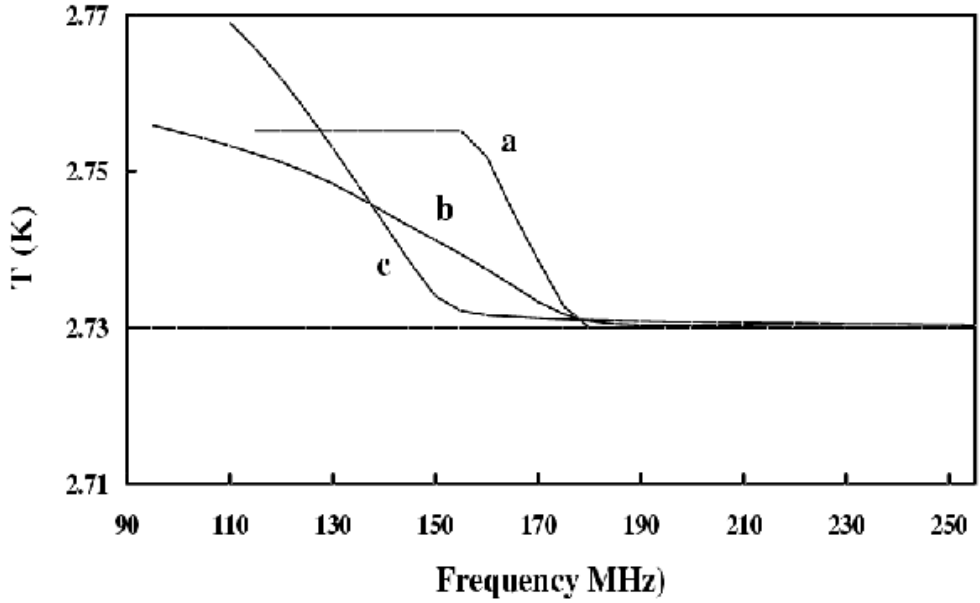}
\caption{The signal expected from the sudden disappearance of HI,
computed as $\delta T_b \simeq (9.0~ {\tt mK}) h^{-1} (\Omega_B h^2/
0.02) [(1+z)/10]^{1/2}$ (from Shaver et al. 1999).}
\end{figure}

\begin{figure}
\centering
\psfig{figure=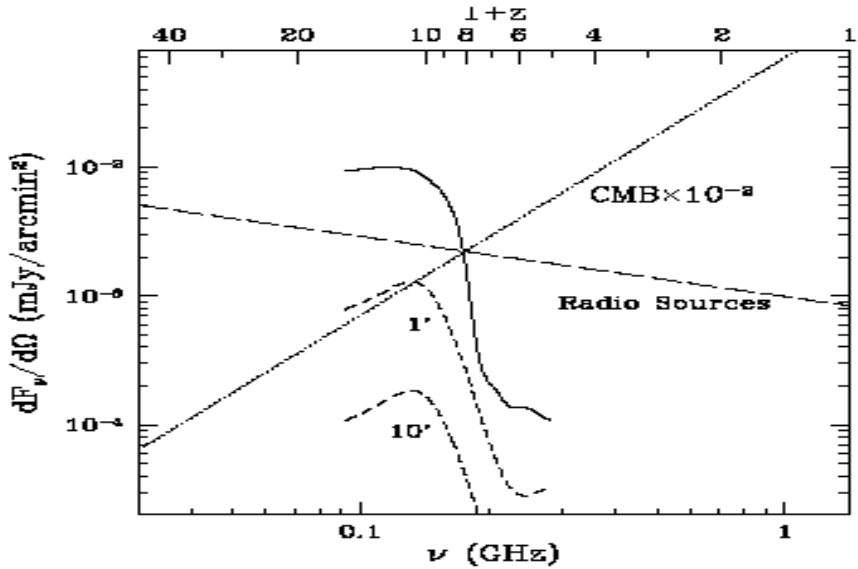,height=8truecm,width=13truecm}
\caption{The signal expected from the sudden disappearance of HI,
including fluctuations on 1' and 10' scale, computed in a $\Lambda$CDM
simulation (from Gnedin \& Ostriker 1997) and compared with the
spectrum of CMB and foreground sources. }
\end{figure}

Another all--sky, distinctive signature is expected at lower
fequencies.  A narrow but deep absorption feature would appear in
correspondence of the short but significant interval when the
Ly$\alpha$ field reach the thermalization value while $T_S<T_K$.  The
signal computed in the case of heating only by Ly$\alpha$ photons, is
shown in Figure 3.  Such a strong absorption feature ($\simeq -40 $ mK)
marks the transition from a cold and dark universe to a universe
populated with radiation sources, but not yet reionized.  The effect
is weakly dependent on the reheating epoch and on the adopted
cosmology.  On the other hand, the amplitude of the signal will be
strongly dependent on the timescale $\tau$ on which the Ly$\alpha$
field reaches the thermalization value.  In any case, the maximum of
the absorption is always larger than 10 mK (see Tozzi et al. 1999).
 
Thus, the reheating and reionization epochs are expected to leave a
double signature superimposed on the CMB in the 100--200 MHz frequency
range.  The detailed appearance of such a feature will depend on the
spectrum of the ionizing sources.  A physical scenario, including a
well defined population of sources, can be built with the code CUBA
(F. Haardt and P. Madau, see {\tt
http://pitto.mib.infn.it/~haardt/cosmology.html}), which provide the
evolution of the background radiation to be plugged into Eq. 2.

\begin{figure}
\psfig{figure=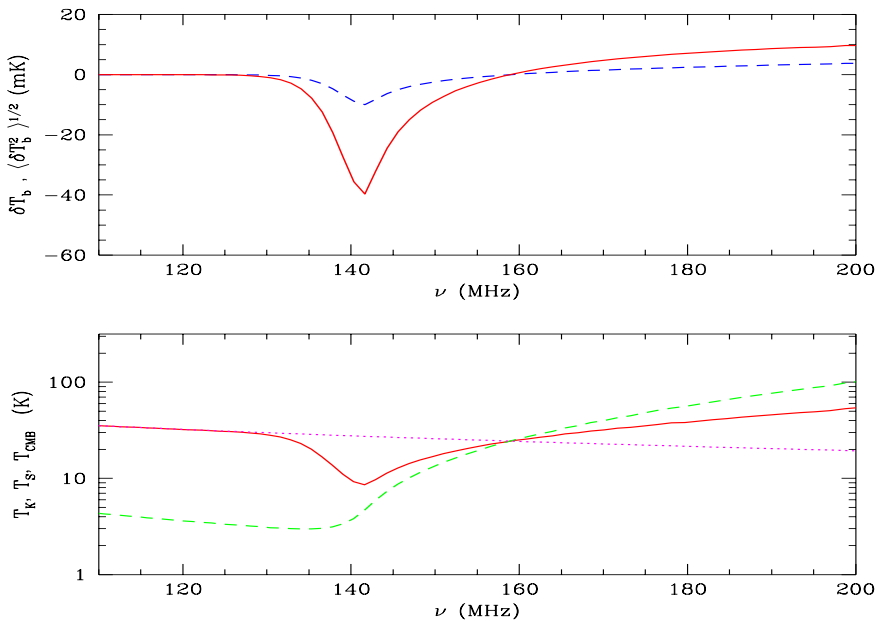,height=9truecm,width=12truecm}
\caption{Top: $\delta T_b$ for a resolution of 1' and 1 MHz in
frequency, both for the fluctuations (dashed line) and for the
continuous distribution of the IGM (solid line), assuming
$P_{\alpha}=P_{th}$ at $z_{th}\simeq 9$ in a critical universe.
Bottom: the corresponding $T_S$ (solid) and $T_K$ (dashed) are
shown together with $T_{CMB}$ (dotted line).}
\end{figure}

A tomography of the IGM at high $z$ can be obtained scanning this
frequency range.  Apart from the all--sky, global signatures that are
potentially detectable with a single--dish telescope, an array with a
sensitivity of few $\mu$Jy can detect also the fluctuations in the
neutral baryons on scales $> 1$', corresponding to few comoving
Mpc.  In the linear regime, the fluctuations induced in the brightness
temperature will be directly proportional to $\Delta \rho /\rho$,
allowing a straightforward reconstruction of the perturbation field at
that epoch.  In Figure~4 we show results for two cosmologies, a tilted
critical CDM universe (tCDM), and an open $\Omega_0=0.4$
universe (OCDM).  In both cases the fluctuations are normalized to
reproduce the local abundance of clusters of galaxies.  In OCDM the
fluctuations are larger by a factor of 3 since for a given local
normalization at $z=0$, the amplitude of the perturbations at high $z$
is correspondingly larger than that in tCDM.  In both figures the
density field has been evolved with a collisionless N--body simulation
using the Hydra code (Couchman, Thomas, \& Pearce 1995), where the
baryons are assumed to trace the dark matter distribution without any
biasing.

\begin{figure}
\centering
\plottwo{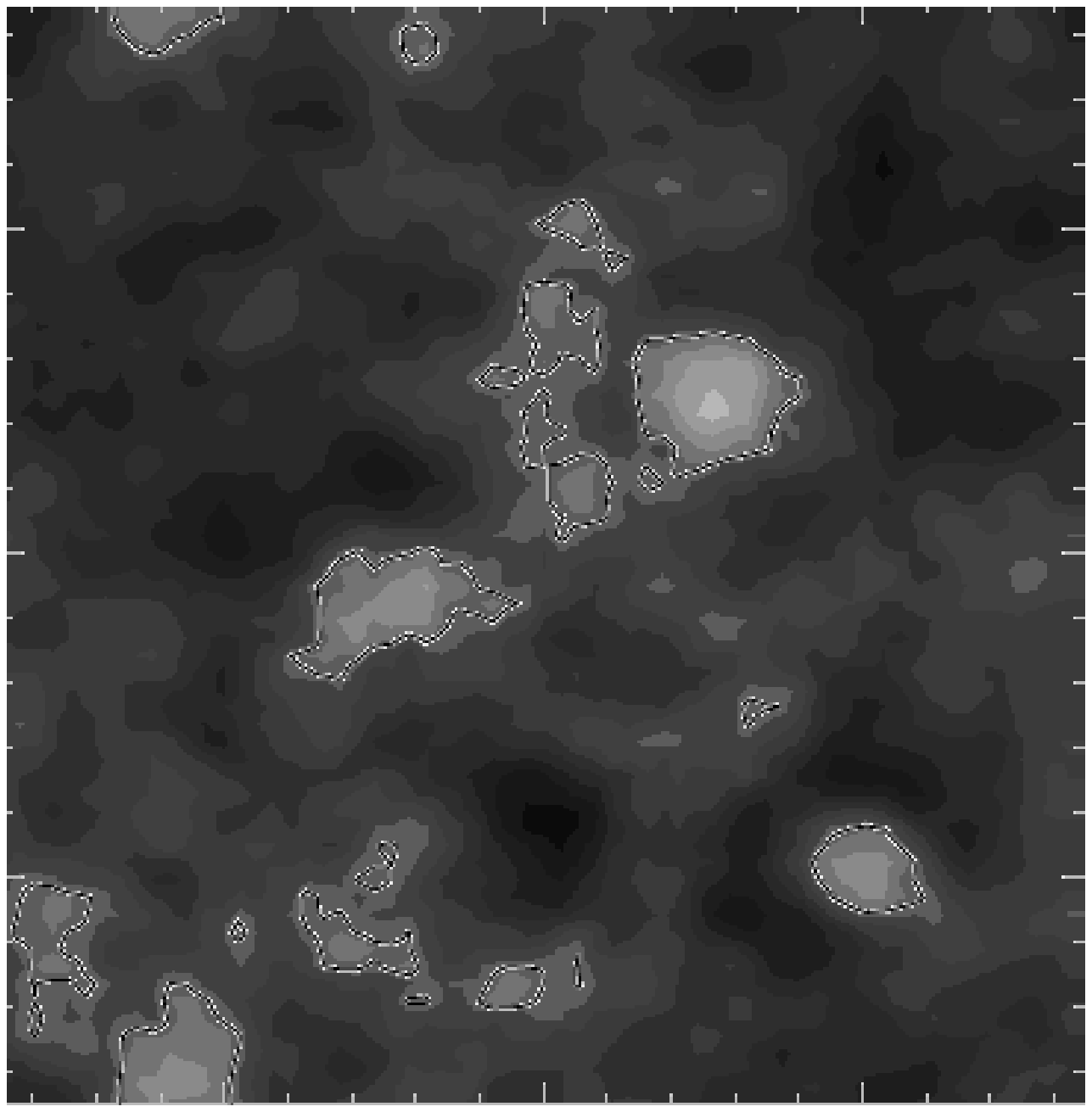}{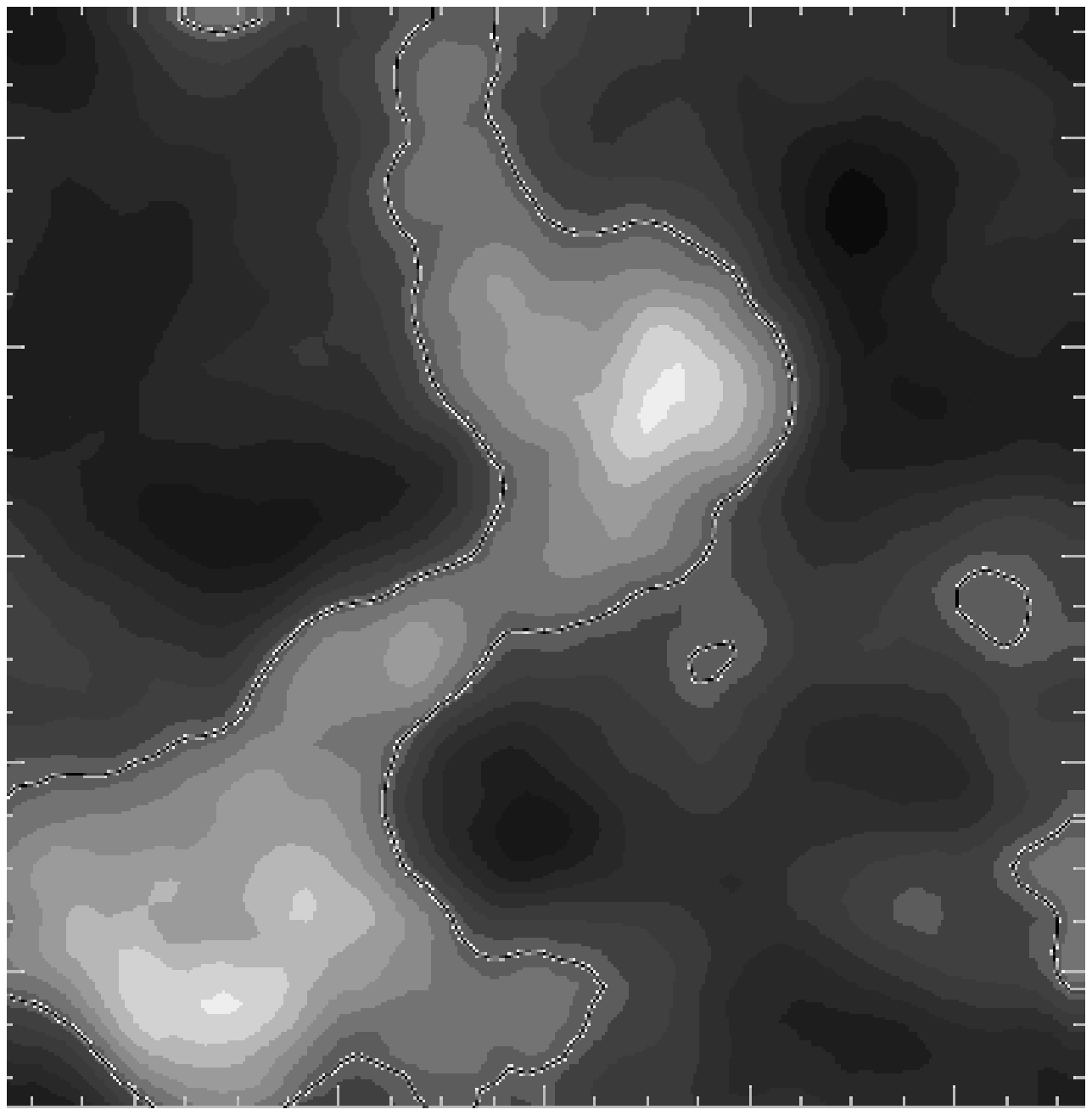}
\caption{Left: Radio map of redshifted 21cm emission against the CMB
in a tCDM cosmology at $z=8.5$ (beam of 2' and bandwidth of 1 MHz).
The linear size of the box is $20h^{-1}$ comoving Mpc, corresponding
to 17' (11') in tCDM (OCDM).  The color intensity goes from $1$
to $6\,$ $\mu$Jy per beam.  The contour levels outline regions with
signal greater than $4\,\mu$Jy per beam. Right: same for OCDM.}
\end{figure}

If reheating is provided by a sparse distribution of quasars, 21cm
emission on Mpc scales will be produced in the quasar neighborhood
(outside the HII bubble) as the medium surrounding it is heated to
$T_S=T_K>T_{\rm CMB}$ by soft {X-rays} from the quasar itself.  The
emission region is followed by an absorption ring, since the
Ly$\alpha$ photons reach regions where $T_K<T_{CMB}$. The size and
intensity of the detectable 21cm region will depend on the quasar
luminosity and age.  The radio map resulting from a quasar `sphere of
influence', 10 Myr after it turns on at $z=8.5$ (tCDM) is shown in
Figure 5 (left).  The absorption region is limited to a small edge.
Moreover, in Figure 5 (right), we show a quasar with the same
Ly$\alpha$ luminosity but with a spectrum absorbed at energies larger
than the Lyman limit, as in the case with an intrinsic $N_H > 10^{22}
$ cm$^{-2}$.  Consequently the HII region is reduced, and the X--ray
warming front is well behind the light radius.  This occurrence leads
to a larger absorption ring where the signal reaches $\simeq -40$ mK.
In addition, imaging the gas surrounding a quasar in 21cm emission
could provide a direct means of measuring intrinsic properties of the
source, like the emitted spectrum and the opening angle of quasar
emission.

\begin{figure}
\plottwo{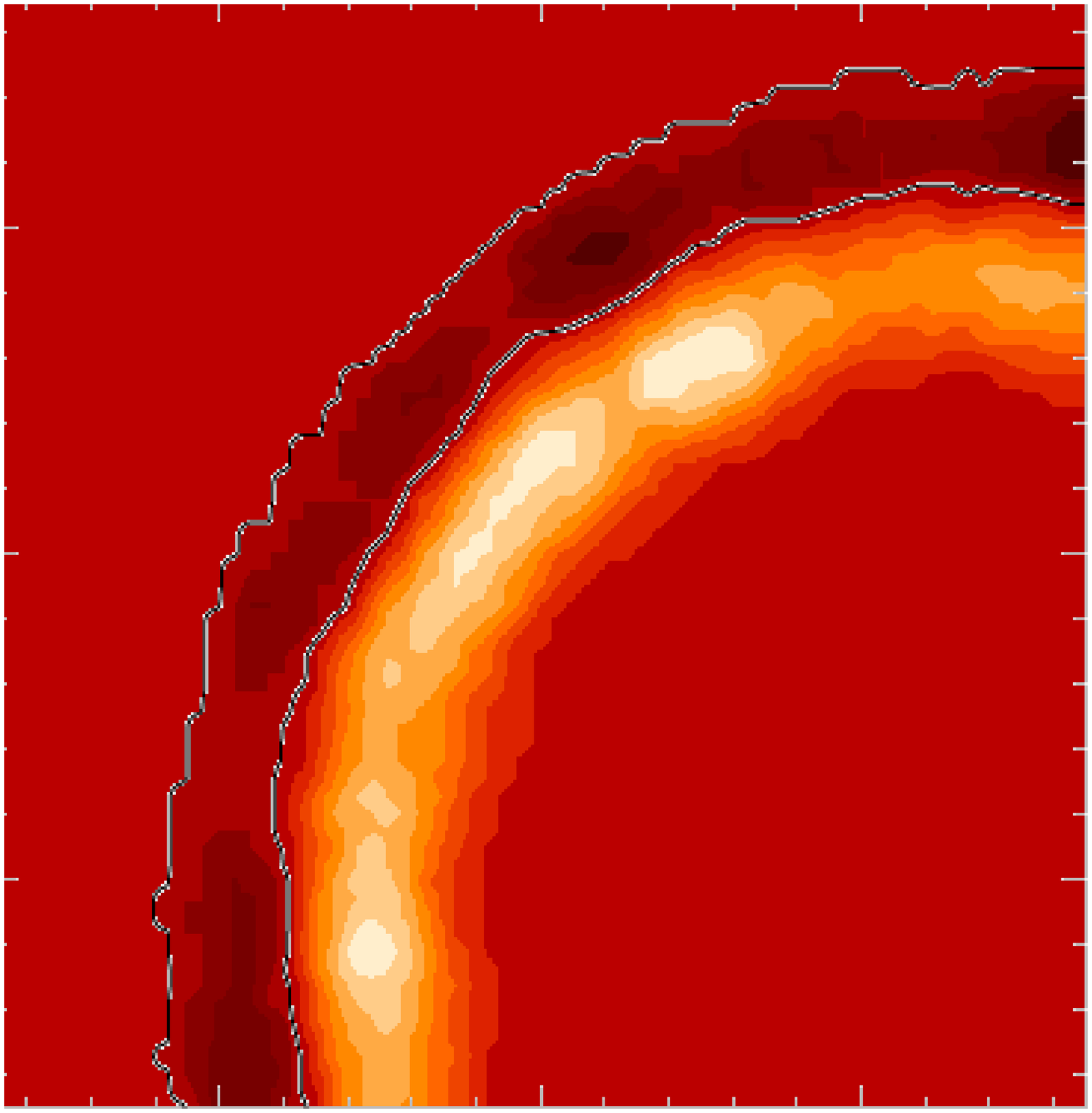}{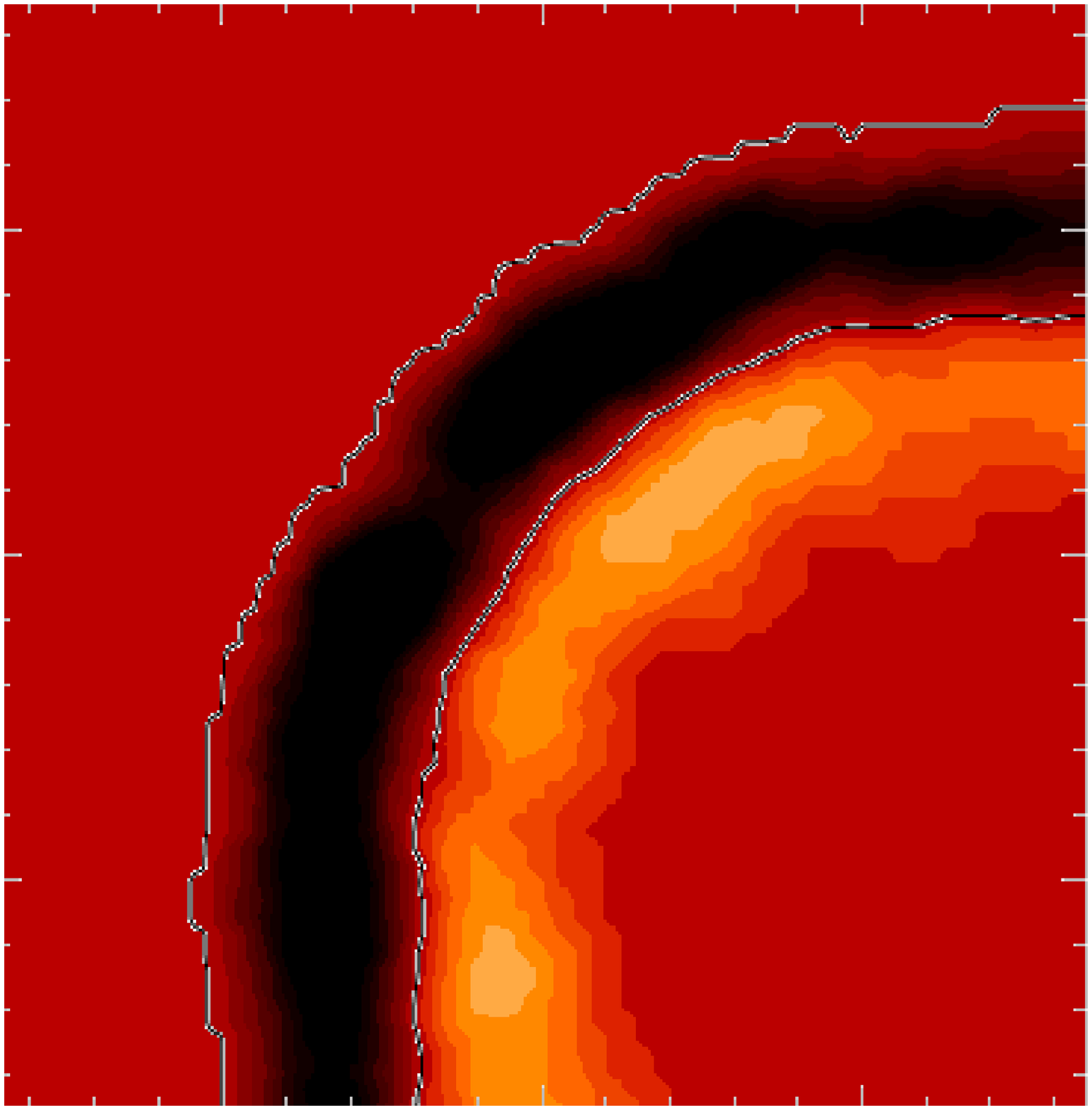}
\caption{Left: 21cm emission and absorption against the CMB from the
region surrounding a quasar.  The size of the box is of $20h^{-1}$
comoving Mpc (tCDM; resolution is 2', frequency depth 1 MHz).  The
color levels range from $-3$ $\mu$Jy to $3$ $\mu$Jy per beam.  The
quasar turns on at $z = 8.5$ with a ionizing luminosity of $10^{57}$
photons s$^{-1}$, and is observed after $10$ Myr.  Right: Same as
figure on the left, but the spectrum of the quasar has an exponential
cutoff at energies larger than the Lyman limit.}
\end{figure}

The proposed observations are an obvious target for the Square
Kilometer Array (SKA, see {\tt
http://www.ras.ucalgary.ca/SKA/science/science.html}).  If $\simeq
2/3$ of SKA were reserved for a compact configuration ($\simeq 7$ km
in diameter) the IGM fluctuations would be detectable at 150 MHz with
1' resolution and a bandwidth of 1 MHz, as also quasar
emission/absorption shells.  In Figure 2 of Tozzi et al. (2000), we
showed that a 5$\sigma$ detection of fluctuations on scales larger
than 2' can be reached with less than 100 hours of integration.

A potential problem of this kind of observation is the extremely low
frequencies involved.  To explore the range $10<z<20$, where the
reheating phase is expected, we need good sensitivity and resolution
down to 70 MHz and lower, where technical problems due to the
structure of the ionosphere arise and a dedicated instrument is
needed.  This bandwidth will be observable with the Low Frequency
Array (LOFAR).  Such observations will have greater probability of
detecting the reheating epoch at high $z$, taking advantage of the
strong signal expected during the short epoch when the cold HI reaches
its maximum optical depth against the CMB.

As we said above, we do not rely only on future radio facilities.
All--sky features or $\simeq 30$' scale fluctuations (corresponding to
$\simeq 1 $ mK) would be detectable with a small but dedicated
telescope, with diameter of about 250 m.  A near--future radio
facility like the Canadian Large Adaptive Reflector (CLAR, see
S. Cot\'e, these Proceedings) would provide a 5$\sigma$ detection in
300 hours of integration for a typical density fluctuation.

\section{Conclusions}

The reionization is a crucial epoch in the evolution of cosmic
structures.  The same ingredients that characterize it (for example:
nature of the first sources, the role of black holes in galaxy
formation, the role of feedback, the chemical enrichment of the IGM)
will deeply affect the evolution of the Universe in the following
epochs. Getting direct information about these crucial early stages of
the luminous Universe is an obvious goal for the next--decade
cosmology.  To date, several strategies have been proposed, but most
of them must wait for the next generation of astronomical facilities.

The observations of high--$z$ HI in the redshifted 21cm, on the other
hand, offer the possibility to investigate the reionization epoch
using future but also present--day radio facilities.  All--sky signals
from HI at redshift $z\simeq 10$ are recognizable as discontinuities
superimposed on the CMB in the wavelength range $\simeq 100-200$ MHz.
In particular, the epoch of reheating can be seen as a deep ($\simeq
-40$ mK) absorption feature against the CMB, at the corresponding
redshifted 21cm line.  The density perturbation field at a redshift
$z\approx 5\div 20$ can be reconstructed looking for mK fluctuations
at $1'$--$5'$ resolution in the radio sky.  Finally, luminous quasars
can be seen by identifying peculiar, ring--shaped signals whose
morphology depends on the source's age, luminosity and geometry.  Such
observations represent an exciting scientific and technological
challenge, and will constitute a unique investigation of the Dark
Ages, for many aspects complementary to future observations at other
wavelengths.

% \acknowledgements 

\end{document}